\documentclass[12pt]{article}%
\usepackage{amssymb}
\usepackage{graphicx}
\usepackage{amsmath}
\usepackage{amsfonts}%
\newtheorem{theorem}{Theorem}

\newtheorem{lemma}[theorem]{Lemma}

\newenvironment{proof}[1][Proof]{\noindent\textbf{#1.} }{\ \rule{0.5em}{0.5em}}
\begin{document}

\title{Poisson Hypothesis for Information Networks II. Cases of Violations and Phase
Transitions\textbf{\ }}
\author{Alexander Rybko\\Institute for the Information Transmission Problems,\\Russian Academy of Sciences, Moscow, Russia,\\rybko@iitp.ru
\and Senya Shlosman\\Centre de Physique Theorique, CNRS,\\Luminy Case 907,\\13288 Marseille, Cedex 9, France\\shlosman@cpt.univ-mrs.fr}
\maketitle

\begin{abstract}
We present examples of queuing networks that never come to equilibrium. That
is achieved by constructing Non-linear Markov Processes, which are
non-ergodic, and possess eternal transience property.

MSC-class: 82C20 (Primary), 60J25 (Secondary)

\end{abstract}

\section{Introduction}

In our previous paper \cite{RS} we have proven the Poisson Hypothesis (PH) for
some simple models of information networks. In a nutshell, PH states that
under general conditions a large network relaxes to a stationary state, in
which different nodes are almost independent, while the relaxation time does
not depend on the size of the network, provided its initial state is properly
chosen. Moreover in this stationary state the incoming flow to every node is
approximately Poissonian with a constant rate, no matter what the service
times are.

The present paper is devoted to the cases of violations of the Poisson
Hypothesis. Our goal is to give examples when the relaxation time for the
network diverges with its size, for certain initial conditions, $\nu$.
Informally speaking, our result means that for certain service time
distributions the network can behave in unstable and unpredictable manner.
More precisely, we know from \cite{RS} that if the expected service time
$S\left(  \nu\right)  $ (see below) of the initial state $\nu$ is finite, then
PH holds. However there exist dangerous states, with infinite expected service
time $S\left(  \nu\right)  $, and they may cause the system to behave
irregularly. This phenomenon can be interpreted as the appearance of the phase
transition in the network.

We will give now the heuristic description of how such an instability can
occur. To be able to construct such an example we need the following property
of the distribution of the service time $\eta$. Supposing that for every
$\tau>0$ the probability that $\eta>\tau$ is positive, we introduce the
\textit{remaining service times}
\[
\eta\Bigm|_{\tau}=\left(  \eta-\tau\Bigm|\eta>\tau\right)  ,\tau\geq0,
\]
and consider the expected values
\[
R_{\eta}\left(  \tau\right)  \equiv\mathbb{E}\left(  \eta\Bigm|_{\tau}\right)
.
\]
In what follows we will suppose that the function $R_{\eta}\left(
\tau\right)  $ \textbf{is unbounded;} otherwise
\[
S\left(  \nu\right)  =\mathbb{E}_{\nu}\left(  R_{\eta}\left(  \tau\right)
\right)
\]
is always finite and PH holds for every initial state, as was proven in
\cite{RS}.\textbf{\ } Let $\mathcal{A}$ be the (big) set of the nodes of our
network. We will consider the closed network -- i.e. customers $\mathcal{C}$
are not leaving the network, and new ones do not come. After being served the
customer goes to some other node and waits for his turn there (see F. Kafka,
Das Schloss). The number of customers, $N=\left\vert \mathcal{C}\right\vert ,$
will be proportional to the number $M=\left\vert \mathcal{A}\right\vert $ of
servers,$\ N=\rho M.$ The service times at different servers are i.i.d. random
variables, with the distribution $\eta.$ Such a model is often used, for
example, in the studies of multiprocessor computer systems, \cite{BS},
\cite{HV}. Suppose for definiteness that $R_{\eta}\left(  0\right)
=\mathbb{E}\left(  \eta\right)  =1$\textit{sec}. Imagine that the initial
state of the network at $t=0$ is such that on a tiny subset $\mathcal{A}%
_{1}\subset\mathcal{A}$ of servers there are customers $\mathcal{C}_{1}$
already under service, $\left\vert \mathcal{C}_{1}\right\vert =\left\vert
\mathcal{A}_{1}\right\vert ,$ and moreover their service lasts already a time
$\tau_{1}\geq1$\textit{min. }If $R_{\eta}\left(  1\text{\textit{min}}\right)
$ is large enough, then it can well happen that almost all the other customers
in the system, $\mathcal{C\setminus C}_{1},$ eventually, after time
$T_{1}^{in}$ will find themselves queuing for the servers in $\mathcal{A}%
_{1},$ and the system will be blocked for a while. Yet after some longer time
$T_{1}^{out}$ the service of the blocking clients $\mathcal{C}_{1}$ will be
over, they leave their nodes, and after still later moment $T_{1}^{bn}$ the
queues at $\mathcal{A}_{1}$ dissolve and network is back to normal, having on
the average $\rho$ clients per server. However, suppose now more: that on a
tiny subset $\mathcal{A}_{2}\subset\mathcal{A}_{1}$ at initial moment $t=0$
the corresponding set $\mathcal{C}_{2}\subset\mathcal{C}_{1}$ of the customers
is in fact served already for a much longer time $\tau_{2}\geq1$\textit{hour,
}while $R_{\eta}\left(  1\text{\textit{hour}}\right)  $ is even larger. Then
the behavior of the network during the time $\left[  T_{1}^{in},T_{1}%
^{out}\right]  $ and $\left[  T_{1}^{out},T_{1}^{bn}\right]  $ will largely
remain the same as above, while after a still later moment $T_{2}^{in}$
essentially all the customers will be queuing for the servers in
$\mathcal{A}_{2}.$ The network again will be blocked during the time interval
$\left[  T_{2}^{in},T_{2}^{out}\right]  ,$ with essentially all the servers
idle, after which the second recovery period, $\left[  T_{2}^{out},T_{2}%
^{bn}\right]  ,$ starts, and after the moment $T_{2}^{bn}$ we have `busyness
as usual' picture, with $\rho$ clients per node on the average. Clearly, this
construction can be iterated, and the number of iterations is bounded only by
the size of the network, $\left\vert \mathcal{A}\right\vert .$ The period of
this oscillatory behavior increases with time.

A similar kind of behavior occurs in other contexts. One such example is the
phenomenon of aging, see e.g. \cite{BDG}.

In this paper we will construct examples of the networks with the above
behavior. Not surprisingly, it turns out to be quite involved technically. The
reason is that the synchronization which we want the system to have, can be
destroyed by fluctuations. Indeed, the long queues, which we want to appear at
the nodes $\mathcal{A}_{1},\mathcal{A}_{2}...$ can be supplemented by other
long queues appearing randomly elsewhere, which will interfere with the
behavior we want. Also, the moments $T_{i}^{out}$ of the beginning of the
queues to dissolve, are not well defined and are in fact not the moments, but
segments of time, their duration growing with $i.$ Therefore the whole picture
desired can be smoothed away in a while.

This is why in the present work we will treat the simplest case of the above
behavior, when the service time $\eta$ assumes only integer values. This
simplification, however, comes with its own cost. The present paper relies
much on our first one, \cite{RS}, which deals only with continuous random
variables $\eta,$ using even the smoothness properties of the density $p,$ as
well as its positiveness everywhere. Therefore our work consists of two parts.
In the first one we construct examples of networks and initial states with the
desired behavior, assuming the validity of the results of \cite{RS} for the
case of integer valued service times. In the two Appendices we present the
extension of \cite{RS} to the integer valued case.

The rest of the paper is organized as follows. In the next section we give the
definition of the non-linear Markov process, the study of which establishes
the validity of properties described above. In Section 3 we introduce
coordinates in one class of these processes. The next section contains the
formulation of our main result and its three subsections present the proof of it.

\textbf{Dynamical systems aspect. }In this paragraph we describe the specific
features of the corresponding dynamical system. As we were explaining in
\cite{RS}, the validity of PH was to some extent a corollary of the statement
that the dynamical system $\mathfrak{D}_{\eta}$, corresponding to the
Non-linear Markov Process, has a global attractor. More precisely, if the
function $R_{\eta}$ is bounded, then $\mathfrak{D}_{\eta}$ has a line $\ell$
of fixed points, and \textbf{any} other point in the phase space is driven by
dynamics to the corresponding point in $\ell,$ which has the same value of the
conserved quantity, $N.$ The results of the present paper give an example of
another feature of $\mathfrak{D}_{\eta}.$ Namely, for some service times
$\eta$, with the function $R_{\eta}$ unbounded, the union of the basins of
attraction of fixed points of the line $\ell$ is not the whole phase space.
Moreover, for some initial points outside this union the corresponding
trajectory $\gamma\left(  t\right)  $ behaves as follows: as $t\rightarrow
\infty,$ the distance $\mathrm{dist}\left(  \gamma\left(  t\right)
,\ell\right)  \rightarrow0$ -- and yet the line $\gamma\left(  t\right)  $ has
no limit as $t\rightarrow\infty,$ and the set of its limit points is a
non-trivial segment of the curve $\ell.$ So in some sense the trajectory
$\gamma\left(  t\right)  $ behaves like the graph of the function $\sin
\frac{1}{x},$ as $x\rightarrow0.$ Such behavior of dynamical system is called
eternal transience, see \cite{M}.

\section{Model description}

Consider the closed network of $M$ servers and $N$ clients, described briefly
in the previous section. Its detailed description is given in the Appendix I.
For large $M,N$ this is a complicated Markov process; but the Weak Poisson
Hypothesis predicts that in the limit $\frac{N}{M}\rightarrow\rho$ it splits
into product of independent random processes, going on different servers. We
prove the Weak Poisson Hypothesis for our setting in the Appendix I, while
here we will describe the limiting processes going on single servers. They
turn out to be Non-linear Markov Processes (NML).

The corresponding dynamical system, describing the evolution of the state
$\mu$ of a single server, $\mathcal{N}$, is determined by one random variable,
$\eta$ -- service time -- whose distribution is given by the probabilities
\[
p\left(  k\right)  =\mathbf{\Pr}\left\{  \eta=k\right\}  ,
\]
with
\[
\sum_{k=1}^{\infty}p\left(  k\right)  =1.
\]
In our application most of $p\left(  k\right)  $ will be zero, but infinitely
many of them will stay positive.

The dynamical system is defined on the space of probability measures $\mu$ on
the set $\Omega=\left\{  \omega=\left(  n,\tau\right)  :n=1,2,...,~\tau
=0,1,2,...\right\}  \cup\left(  0,0\right)  ,$ so $\Omega\subset\mathbb{Z}%
^{2}.$ We will denote by $\mathbf{0}$ the special point $\left(  0,0\right)  .
$ The notation $n\left(  \omega\right)  ,\tau\left(  \omega\right)  $ mean the
first, resp., second coordinate of $\omega.$ Therefore the measure $\mu$ is a
collection of non-negative numbers $\mu\left(  n,\tau\right)  ,$ with
\[
\sum_{n=1,\tau=0}^{\infty}\mu\left(  n,\tau\right)  +\mu\left(  \mathbf{0}%
\right)  =1.
\]

Let the measure $\mu^{T}$ on the $T$-th step is given. We will describe now
the way it evolves in one time-step to the measure $\mu^{T+1}.$ This goes in
three stages.

\textbf{A. Vertical shift. }

We construct the measure $\phi^{T+1}$ by
\begin{equation}
\phi^{T+1}\left(  n,\tau\right)  =\left\{
\begin{array}
[c]{ll}%
\mu^{T}\left(  n\mathbf{,}\tau-1\right)  & \text{ if }\tau>0\\
0 & \text{ for }\tau=0
\end{array}
\right.  , \label{11}%
\end{equation}
\[
\phi^{T+1}\left(  \mathbf{0}\right)  =\mu^{T}\left(  \mathbf{0}\right)  .
\]

\textbf{B. Downward jump. (Exit flow)}

We introduce the conditional probabilities $p_{\tau}$ by
\begin{equation}
p_{\tau}=\frac{p\left(  \tau\right)  }{\sum_{k\geq\tau}p\left(  k\right)  }.
\label{32}%
\end{equation}
We define the measure $\psi^{T+1}$ by
\begin{equation}
\psi^{T+1}\left(  n,\tau\right)  =\left\{
\begin{array}
[c]{ll}%
\phi^{T+1}\left(  n,\tau\right)  \left(  1-p_{\tau}\right)  & \text{if }%
\tau\geq1,\\
\sum_{\tau\geq1}\phi^{T+1}\left(  n+1,\tau\right)  p_{\tau} & \text{if }%
n\geq1,~\tau=0,\\
\phi^{T+1}\left(  \mathbf{0}\right)  +\sum_{\tau\geq1}\phi^{T+1}\left(
1,\tau\right)  p_{\tau} & \text{if }n=\tau=0.
\end{array}
\right.  \label{3}%
\end{equation}

\textbf{C. Right-shift. (Inflow)}

Let $\pi_{\lambda}\left(  k\right)  $ be the distribution of the Poisson
random variable;
\begin{equation}
\pi_{\lambda}\left(  k\right)  =e^{-\lambda}\frac{\lambda^{k}}{k!}. \label{2}%
\end{equation}
We will interpret the function $\pi_{\lambda}\left(  k\right)  $ as a measure
on $\mathbb{Z}^{1}\subset\mathbb{Z}^{2},$ supported by the points $\left(
k,0\right)  .$ Likewise, we think about all the measures introduced above, as
measures on $\mathbb{Z}^{2}.$ Let
\begin{equation}
\lambda\left(  T+1\right)  =\sum_{n\geq1}\sum_{\tau\geq1}\phi^{T+1}\left(
n,\tau\right)  p_{\tau} \label{4}%
\end{equation}
be the total mass which went down (to the level $\tau=0,$ see $\left(
\ref{3}\right)  $) on the previous step. Note for the future that
\begin{equation}
\lambda\left(  T+1\right)  \leq1. \label{25}%
\end{equation}

We put finally
\begin{equation}
\mu^{T+1}\left(  n,\tau\right)  =\left[  \psi^{T+1}\ast\pi_{\lambda\left(
T+1\right)  }\right]  \left(  n,\tau\right)  \equiv\sum_{k}\psi^{T+1}\left(
n-k,\tau\right)  \pi_{\lambda\left(  T+1\right)  }\left(  k\right)  .
\label{12}%
\end{equation}

\textbf{Particle conservation property. }Note that by construction the mean
value
\[
N\left(  \mu^{T}\right)  \equiv\sum_{\omega\in\Omega}n\left(  \omega\right)
\mu^{T}\left(  \omega\right)
\]
does not change with $T.$

\section{Choice of the distribution of $\eta$ and of the initial state $\nu$}

Let $A=\left\{  a_{1}=1<a_{2}<a_{3}<...\right\}  \subset\mathbb{N\cup}\left\{
\infty\right\}  $ be an infinite subset. Then the probabilities $p\left(
k\right)  $ will be chosen to vanish unless $k=a_{i}$ for some $i.$

The probabilities $p\left(  a_{n}\right)  $ are given by
\begin{equation}
p\left(  a_{n}\right)  =\sum_{a_{n}\leq l<a_{n+1}}\left(  \frac{1}{2}\right)
^{l}. \label{13}%
\end{equation}
We denote the conditional probabilities $p_{a_{n}}$ by $q_{n};$%
\[
q_{n}=\frac{p\left(  a_{n}\right)  }{\sum_{k\geq a_{n}}p\left(  k\right)  }.
\]
We want the conditional probabilities $q_{n}$ to increase to $1$ very fast.
This can be ensured by taking the set $A$ very sparse.

In what follows we will use the convention that both relations: $\infty
<\infty$ and $\infty\leq\infty$ -- are valid.

We parametrize the sets $A$ in the following way. Let the integers
$C_{0}=0<B_{1}=1\leq C_{1}<B_{2}\leq C_{2}<...\leq\infty$ be given. We say
that the set $A$ is of type $\mathcal{B}=\left\{  B_{1},C_{1},B_{2}%
,C_{2},...\right\}  ,$ if $A\subset\left\{  \cup_{k=1}^{\infty}\left[
B_{k},C_{k}\right]  \right\}  ,$ and each intersection $A\cap\left[
B_{k},C_{k}\right]  \neq\varnothing.$

By definition, the initial state $\nu$ will be supported by the set
\begin{equation}
\mathcal{D}=\left\{  \left(  n,\tau\right)  :1\leq n\leq\bar{n},\tau\in
\cup_{k=0}^{\infty}\left[  C_{k},C_{k}+F_{k+1}\right]  \right\}  \subset
\Omega, \label{31}%
\end{equation}
where $\bar{n}\geq1$ is fixed, as well as the integers $F_{1}=0\leq
F_{k}<\infty$ . We will assume that $C_{k}+F_{k+1}<B_{k+1}$ for all $k\geq0.$
We further assume that the differences $G_{k}=C_{k}-B_{k}\geq0$ will be finite
and fixed. It will be convenient for us to parametrize the family of $\nu$-s
we will consider as follows. For every $k\geq0$ let us fix in an arbitrary way
a probability measure $\varkappa_{k}$ on the finite set $\left\{  \left(
n,\tau\right)  :1\leq n\leq\bar{n},\tau\in\left[  C_{k},C_{k}+F_{k+1}\right]
\right\}  .$ (For example, we can take all $\varkappa_{k}$ to be $\delta
$-measures having a unit atom at the point $\left(  1,C_{k}\right)  .$) Now
for every collection
\[
\Delta=\left\{  d_{k}\geq0\right\}  ,\ \sum_{k:B_{k}<\infty}d_{k}=1
\]
we define the probability measure $\nu_{\Delta}$ on $\Omega$ by
\begin{equation}
\nu_{\Delta}=\sum_{k}d_{k}\varkappa_{k}. \label{14}%
\end{equation}
In what follows, the choice of the integer $\bar{n},$ integers $F_{k},$
integers $G_{k}=C_{k}-B_{k}$ and the measures $\varkappa_{k}$ will be fixed.
It will be the choice of the parameters $B_{k}$ and $\Delta$ that we will have
to make. We will define them inductively. This will be possible in our case
because by construction which follows, for every time duration $T,$ however
large, the relevant features of our dynamical system will depend only on the
first few values $B_{k},d_{k},$ $k\leq k\left(  T\right)  ;$ though of course
$k\left(  T\right)  \rightarrow\infty$ as $T\rightarrow\infty.$

In the following we will need that the distances $B_{k+1}-\left(
C_{k}+F_{k+1}\right)  $ grow fast enough. Namely, we will need that for all $k
$%
\begin{equation}
B_{k+1}-\left(  C_{k}+F_{k+1}\right)  >B_{k}. \label{29}%
\end{equation}

We will denote the server thus constructed, with the initial state specified,
as $\mathcal{N}\left(  \mathcal{B},\Delta\right)  .$ To formulate our main
result we need to introduce the mean service time $E=\mathbb{E}\left(
\eta\right)  $ (which is a function of $\mathcal{B}$) and the average number
of clients $\rho=N\left(  \nu_{\Delta}\right)  .$

\section{The main result}

\begin{theorem}
There exists a choice of the parameters $\mathcal{B}$ and $\Delta,$ such that
the corresponding input rate function $\lambda_{\mathcal{B},\Delta}\left(
t\right)  $ ($=$output rate function), see $\left(  \ref{4}\right)  $ -- has
no limit as $t\rightarrow\infty.$ Moreover, given $\varepsilon>0,$ there
exists a choice of $\mathcal{B},\Delta$ such that

\begin{itemize}
\item there exists a sequence of disjoint segments of time moments,
$t\in\left[  T_{k}^{in},T_{k}^{out}\right]  ,$ $\lim_{k\rightarrow\infty}%
T_{k}^{in\left(  out\right)  }\rightarrow\infty,$ where $\lambda
_{\mathcal{B},\Delta}\left(  t\right)  \leq\varepsilon$.

\item there exists an increasing sequence of time moments $T_{k}%
^{bn}\rightarrow\infty,$ such that
\begin{equation}
\lambda_{\mathcal{B},\Delta}\left(  T_{k}^{bn}\right)  \geq c_{PK}\left(
\eta,\rho\right)  -\varepsilon, \label{0002}%
\end{equation}
where the rate $c_{PK}\left(  \eta,\rho\right)  >0$ is given by
Pollaczek-Khinchin relation.

\noindent As a result, the state $\mu^{t}$ has no limit as $t\rightarrow
\infty$.
\end{itemize}
\end{theorem}

The main idea of studying the evolution of the server $\mathcal{N}\left(
\mathcal{B},\Delta\right)  $ is to treat it as a small perturbation of the
servers $\mathcal{N}_{n}$ having the form $\mathcal{N}\left(  \mathcal{B}%
_{n},\tilde{\Delta}_{n}\right)  ,$ where the sequence $\mathcal{B}%
_{n}=\left\{  B_{1},C_{1},B_{2},C_{2},...\right\}  $ has the first $2n$
entries the same as in $\mathcal{B},$ while $B_{k}=\infty$ once $k\geq n.$ The
sequence $\tilde{\Delta}_{n}$ has the first $n-1$ coordinates being the same
as in $\Delta,$ the $n$-th one is equal to $\sum_{k\geq n}d_{k},$ and the rest
of them are zeroes. The corresponding service time $\eta_{k}$ thus assumes
only finitely many values, and the vector $\tilde{\Delta}_{n}$ has at most $n$
non-zero entries. Of course, the evolution of $\mathcal{N}\left(
\mathcal{B},\Delta\right)  $ can be close to the one of $\mathcal{N}\left(
\mathcal{B}_{n},\tilde{\Delta}_{n}\right)  $ only for finite duration of time,
and for longer times we will need to go to a finer approximation, with bigger
value of the cutoff $n.$ We will define the parameters $\mathcal{B}_{n}%
,\tilde{\Delta}_{n}$ by induction, their definitions being a function of the
dynamics of the servers with the lower cutoffs.

We will use the following notation:
\begin{equation}
\Omega_{\left[  a,b\right]  }=\left\{  \omega=\left(  n,\tau\right)
:a\leq\tau\leq b\right\}  . \label{22}%
\end{equation}
In particular, $\Omega_{k}=\left\{  \omega=\left(  n,\tau\right)
:\tau=k\right\}  .$ We put
\begin{equation}
N_{k}\left(  \mu\right)  \equiv\sum_{\omega\in\Omega_{k}}n\left(
\omega\right)  \mu\left(  \omega\right)  , \label{1}%
\end{equation}
\begin{equation}
N_{\left[  a,b\right]  }\left(  \mu\right)  \equiv\sum_{\omega\in
\Omega_{\left[  a,b\right]  }}n\left(  \omega\right)  \mu\left(
\omega\right)  . \label{23}%
\end{equation}

\subsection{The first step of induction}

On the first step we will compare the server $\mathcal{N}\left(
\mathcal{B},\Delta\right)  $ with the server $\mathcal{N}_{1}=\mathcal{N}%
\left(  \mathcal{B}_{1},\tilde{\Delta}_{1}\right)  .$ $\mathcal{B}_{1}$ is
just the pair $\left\{  B_{1}=1,C_{1}\right\}  ,$ the service time $\eta_{1}$
takes (integer) values in the segment $\left[  1,C_{1}\right]  ,$ (taking
value $1$ with the probability at least $\frac{1}{2},$ see $\left(
\ref{13}\right)  $), and the initial state is the measure $\varkappa_{1},$
supported by the segment $\left\{  \left(  n,0\right)  :1\leq n\leq\bar
{n}\right\}  .$ The sequence $\tilde{\Delta}_{1}$ has one non-zero component,
$d_{1},$ which equals to $1.$ The result of this comparison will allow us to
determine the values of the parameters $B_{2}$ and $d_{1}$ -- and therefore
$C_{2}$ -- of the server $\mathcal{N}\left(  \mathcal{B},\Delta\right)  .$

The evolution $\mu_{1}^{T}$ (see $\left(  \ref{11}\right)  $--$\left(
\ref{12}\right)  $) of our server with the service times $\left(
\ref{13}\right)  $ and initial state $\left(  \ref{14}\right)  $, is
relatively simple.

{\small For the benefit of the reader let us describe the server }%
$\mathcal{N}_{1}=N\left(  \mathcal{B}_{1},\tilde{\Delta}_{1}\right)
${\small \ in a special case when }$C_{1}=B_{1}=1.${\small \ We denote it by
}$\mathcal{N}_{1}^{\ast}.${\small \ Then the service time }$\eta${\small \ is
non-random: }$\eta=1${\small \ with probability one. All the measures }%
$\mu_{1}^{T}${\small \ are supported by }$\left\{  \mathbf{0}\right\}
\cup\mathbb{N}.${\small \ The initial state }$\mu_{1}^{0}${\small \ is the
measure }$\varkappa_{1}.${\small \ Denote by }$\rho_{1}=\mathbb{E}\left(
\varkappa_{1}\right)  .${\small \ Let }$\mu_{1}^{T}${\small \ be the measure
after }$T${\small \ steps. Let us write it as }
\[
\mu_{1}^{T}=c\left(  T\right)  \delta_{\mathbf{0}}+\left(  1-c\left(
T\right)  \right)  \alpha^{T},
\]
{\small where }$\delta_{\mathbf{0}}${\small \ is an unit atom at }%
$0,${\small \ while }$\alpha^{T}${\small \ has support in }$\mathbb{N}%
.${\small \ Then we have }
\begin{equation}
\psi^{T+1}\left(  n,0\right)  =c\left(  T\right)  \delta_{\mathbf{0}}\left(
n\right)  +\left(  1-c\left(  T\right)  \right)  \alpha^{T}\left(  n+1\right)
\label{5}%
\end{equation}
{\small (see }$\left(  \ref{3}\right)  ${\small ), }
\begin{equation}
\lambda\left(  T+1\right)  =\alpha^{T}\left(  \mathbb{N}\right)  \label{6}%
\end{equation}
{\small (see }$\left(  \ref{4}\right)  ${\small ), and finally }
\begin{equation}
\mu_{1}^{T+1}=\psi^{T+1}\ast\pi_{\lambda\left(  T+1\right)  } \label{7}%
\end{equation}
{\small (see }$\left(  \ref{12}\right)  ${\small ).}

{\small Another way of describing it is the following: let }$\xi^{T}%
${\small \ be the integer random variable, having distribution }$\mu_{1}^{T}.
${\small \ Define }
\[
\zeta^{T}=\zeta^{T}\left(  \xi^{T}\right)  =\left\{
\begin{array}
[c]{ll}%
0 & \text{ if }\xi^{T}=0,\\
\xi^{T}-1 & \text{ if }\xi^{T}>0.
\end{array}
\right.
\]
{\small Then }$\xi^{T+1}=\zeta^{T}+\gamma^{T+1},${\small \ where }%
$\gamma^{T+1}${\small \ is the Poisson random variable, independent of }%
$\zeta^{T}, ${\small \ with the distribution }$\pi_{\mathbf{\Pr}\left\{
\xi^{T}>0\right\}  }.${\small \ The conservation law }$\mathbb{E}\left(
\xi^{T}\right)  =\mathbb{E}\left(  \xi^{T+1}\right)  =\rho_{1}${\small \ is
immediate.}

The Poisson Hypothesis for the server $\mathcal{N}_{1}=\mathcal{N}\left(
\mathcal{B}_{1},\tilde{\Delta}_{1}\right)  $ implies the weak convergence:
$\mu_{1}^{T}\rightarrow\nu_{1}^{\rho_{1}}$ as $T\rightarrow\infty,$ where
$\nu_{1}^{\rho_{1}}$ is the unique measure on $\Omega_{\left[  0,C_{1}%
-1\right]  },$ which is invariant under the above dynamics and which satisfies
$N\left(  \nu_{1}^{\rho_{1}}\right)  =N\left(  \varkappa_{1}\right)
\equiv\rho_{1}.$ The rate function $\lambda_{\mathcal{B}_{1},\tilde{\Delta
}_{1}}\left(  t\right)  $ goes to the limit value, $\Lambda_{\mathcal{B}%
_{1},\tilde{\Delta}_{1}}.$ The PH validity follows from \cite{RS} and the
Appendix of the present paper.

{\small In the special case of the server }$\mathcal{N}_{1}^{\ast}%
${\small \ the Poisson Hypothesis was obtained earlier in \cite{S}. The
limiting value }$\Lambda_{1}${\small \ of the rate function }$\lambda
_{1}\left(  T\right)  ${\small \ as }$T\rightarrow\infty${\small \ in that
case equals to }$\nu_{1}^{\rho_{1}}\left(  \mathbb{N}\right)  =1-\nu_{1}%
^{\rho_{1}}\left(  \mathbf{0}\right)  .${\small \ }

The convergence $\mu_{1}^{T}\rightarrow\nu_{1}^{\rho_{1}}$ can be expressed in
words: \textquotedblleft the measures $\mu_{1}^{T}$ become \textit{very close}
to the measure $\nu_{1}^{\rho_{1}}$\textquotedblright, if we will use the term
\textquotedblleft\textit{very close to }$\nu_{1}^{\rho_{1}}$\textquotedblright%
\ to mean the following. Denote by $I_{1}\subset\Omega_{\left[  0,C_{1}%
-1\right]  }$ the smallest rectangle of the type $\left\{  \omega=\left(
n,\tau\right)  :n\leq n_{1},\tau\in\left[  0,C_{1}-1\right]  \right\}  $,
having the property
\begin{equation}
N\left(  \nu_{1}^{\rho_{1}}\right)  -\sum_{\left(  n,\tau\right)  \in I_{1}%
}n\nu_{1}^{\rho_{1}}\left(  n,\tau\right)  \leq\frac{1}{100}. \label{30}%
\end{equation}
We say that $\mu$ is $\varepsilon$-close to $\nu_{1}^{\rho_{1}}$ iff for every
$\omega\in I_{1}$ we have
\[
1-\varepsilon<\frac{\mu\left(  \omega\right)  }{\nu_{1}^{\rho_{1}}\left(
\omega\right)  }<1+\varepsilon,\ \ 1-\varepsilon<\frac{\nu_{1}^{\rho_{1}%
}\left(  \omega\right)  }{\mu\left(  \omega\right)  }<1+\varepsilon.
\]
In the following the number $\varepsilon>0$ will be some small fixed quantity.
We define now the time $T_{1}^{bn}$ as the first moment $T$ when the measure
$\mu_{1}^{T}$ becomes $\varepsilon$-close to $\nu_{1}^{\rho_{1}}.$

Note now that if $I\subset\Omega_{\left[  0,C_{1}-1\right]  }$ is some finite
rectangle, and $T$ is some finite time moment, then the restriction $\mu
^{T}\Bigm|_{I}$ of the state of the server $\mathcal{N}\left(  \mathcal{B}%
,\Delta\right)  $ depends on the parameters $\mathcal{B=}\left\{  B_{1}%
,C_{1},B_{2},C_{2},...\right\}  ,$ $\Delta=\left\{  d_{k}\geq0\right\}  $ of
the server $\mathcal{N}$ continuously. We will apply this continuity at the
point $\mathcal{B}_{1},\tilde{\Delta}_{1}.$ We can then claim that there exist
a value $\check{B}_{2}<\infty$ and a value $\hat{d}_{1}<1$ such that for any
consistent choice of the sets $\mathcal{B}$ and $\Delta$ of the parameters
$B_{2}<B_{3}<...$ and $d_{1},d_{2},...:$ $d_{k}\geq0,\ \sum d_{k}=1,$
satisfying
\begin{equation}
B_{2}\geq\check{B}_{2}\text{ and }1>d_{1}\geq\hat{d}_{1}, \label{28}%
\end{equation}
the state $\mu^{T_{1}^{bn}}$ of the corresponding server $\mathcal{N}\left(
\mathcal{B},\Delta\right)  $ is $2\varepsilon$-close to $\nu_{1}^{\rho_{1}}.$
That means in particular that any such server at the moment $T_{1}^{bn}$ has
the property that its input rate $\lambda_{\mathcal{B},\Delta}\left(
T_{1}^{bn}\right)  $ differs very little from the rate $\lambda_{\mathcal{B}%
_{1},\tilde{\Delta}_{1}}\left(  T_{1}^{bn}\right)  ,$ which in turn is close
to its limiting value $\Lambda_{\mathcal{B}_{1},\tilde{\Delta}_{1}}.$ In the
following we will assume the relation $\left(  \ref{28}\right)  $ to be valid
for all servers $\mathcal{N}\left(  \mathcal{B},\Delta\right)  $ further considered.

The choice of the value of the parameter $d_{1}\in\Delta$ can now be made: we
can take it to be equal to $\hat{d}_{1}.$ However we have to work more to be
able to choose $B_{2}\geq\check{B}_{2}$.

To do this we will look on the quantities
\[
N_{\left[  0,C_{1}-1\right]  }\left(  \mu\right)  \equiv\sum_{\omega\in
\Omega_{\left[  0,C_{1}-1\right]  }}n\left(  \omega\right)  \mu\left(
\omega\right)  ,
\]
where $\Omega_{\left[  0,C_{1}-1\right]  }=\left\{  \left(  n,\tau\right)
:\tau\in\left[  0,C_{1}-1\right]  \right\}  ,$ see $\left(  \ref{1}\right)  .$
Clearly, the value $N_{\left[  0,C_{1}-1\right]  }\left(  \mu_{1}^{T}\right)
$ does not depend on $T,$ since $N_{\left[  0,C_{1}-1\right]  }\left(  \mu
_{1}^{T}\right)  =N\left(  \mu_{1}^{T}\right)  ,$ while the latter is
conserved quantity. The quantity $N\left(  \mu^{T}\right)  ,$ corresponding to
the server $\mathcal{N}\left(  \mathcal{B},\Delta\right)  ,$ is also
conserved, but not the $N_{\left[  0,C_{1}-1\right]  }\left(  \mu^{T}\right)
;$ moreover, we are going to show that for every $\varepsilon>0$ we can find
the time duration $T=T\left(  \varepsilon\right)  ,$ such that if $B_{2}>T$
then the function $N_{\left[  0,C_{1}-1\right]  }\left(  \mu^{t}\right)  $
decays with $t$ for $1\leq t\leq T$ and becomes less than $\varepsilon$ at
$t=T\left(  \varepsilon\right)  .$ We start with a simple computation.

Let $\mu$ be a positive measure on $\left\{  \mathbf{0}\right\}
\cup\mathbb{N},$ with $\mu\left(  \left\{  \mathbf{0}\right\}  \cup
\mathbb{N}\right)  =1-\delta,$ $\delta>0.$ Let $\alpha$ be a positive measure
on $\mathbb{N}, $ such that $\alpha\leq\mu;$ here it means that for every
$k\geq0$ $~\alpha\left(  k\right)  \leq\mu\left(  k\right)  .$ Denote by $m$
the value $\alpha\left(  \mathbb{N}\right)  .$ Define $\bar{\mu}$ by
\[
\bar{\mu}\left(  n\right)  =\left\{
\begin{array}
[c]{ll}%
\mu\left(  n\right)  -\alpha\left(  n\right)  +\alpha\left(  n+1\right)  &
\text{ if }n>0,\\
\mu\left(  0\right)  +\alpha\left(  1\right)  & \text{ if }n=0.
\end{array}
\right.
\]
Finally, let $\pi_{m}$ be a probability measure on $\left\{  \mathbf{0}%
\right\}  \cup\mathbb{N},$ with $\mathbb{E}\left(  \pi_{m}\right)  =m,$ and we
define $\mu^{+1}=\bar{\mu}\ast\pi_{m}$ (compare with the dynamics $\left(
\ref{5}\right)  $--$\left(  \ref{7}\right)  $).

\begin{lemma}
\label{decay}
\begin{equation}
N_{0}\left(  \mu\right)  -N_{0}\left(  \mu^{+1}\right)  =m\delta. \label{21}%
\end{equation}

\end{lemma}

Note that if $\mu$ is itself a probability measure, then $\delta=0$ and under
the above transformation the mean value $N_{0}\left(  \mu\right)  $ is
conserved, as it should be.

\begin{proof}
The proof consists of a straightforward computation.
\[
N_{0}\left(  \mu^{+1}\right)  =N_{0}\left(  \bar{\mu}\ast\pi_{m}\right)
=\left(  1-\delta\right)  \mathbb{E}\left[  \left(  \frac{1}{1-\delta}\bar
{\mu}\right)  \ast\pi_{m}\right]
\]
\[
=N_{0}\left(  \bar{\mu}\right)  +\left(  1-\delta\right)  m.
\]
Therefore
\[
N_{0}\left(  \mu\right)  -N_{0}\left(  \mu^{+1}\right)  =N_{0}\left(
\mu\right)  -N_{0}\left(  \bar{\mu}\right)  -\left(  1-\delta\right)  m
\]
\[
=\sum_{k\geq1}k\alpha\left(  k\right)  -\sum_{k\geq1}k\alpha\left(
k+1\right)  -\left(  1-\delta\right)  m
\]
\[
=\sum_{k\geq1}k\alpha\left(  k\right)  -\sum_{k\geq1}\left(  k+1\right)
\alpha\left(  k+1\right)  +\sum_{k\geq1}\alpha\left(  k+1\right)  -\left(
1-\delta\right)  m
\]
\[
=\sum_{k\geq1}\alpha\left(  k\right)  -\left(  1-\delta\right)  m=\delta m.
\]

\end{proof}

Applying the above Lemma to the evolution $\mu^{t}$ of the server
$\mathcal{N}\left(  \mathcal{B},\Delta\right)  $ we have that for any
$t<B_{2}-\left(  C_{1}+E_{2}\right)  $%
\[
N_{0}\left(  \mu^{t}\right)  -N_{0}\left(  \mu^{t+1}\right)  \geq\left(
1-\hat{d}_{1}\right)  \lambda_{\left(  \mathcal{B},\Delta\right)  }\left(
t\right)  .
\]
Here we have the inequality, rather than equality of $\left(  \ref{21}\right)
$, because, in fact, once $B_{2}<\infty,$ the measure $\mu^{t}\left(
\Omega_{\left[  0,C_{1}-1\right]  }\right)  $ decays with $t,$ and so $\mu
^{t}\left(  \Omega_{\left[  0,C_{1}-1\right]  }\right)  \leq\mu^{t=0}\left(
\Omega_{\left[  0,C_{1}-1\right]  }\right)  =\hat{d}_{1}.$ Note that the
applicability of the Lemma is based on relation $\left(  \ref{29}\right)  .$
Since the total mean number of customers, $N\left(  \mu^{t}\right)  ,$ is a
conserved quantity, that can happen only if the rate $\lambda_{\left(
\mathcal{B},\Delta\right)  }\left(  t\right)  $ takes arbitrarily small values
for arbitrarily large $t$-s (not exceeding the threshold $B_{2}-\left(
C_{1}+E_{2}\right)  $). We now choose the value $\hat{B}_{2}$ as the one satisfying:

\begin{enumerate}
\item $\hat{B}_{2}\geq\check{B}_{2},$

\item $\hat{B}_{2}-\left(  C_{1}+E_{2}\right)  >T_{1}^{bn},$

\item there exists a time moment $t\in\left[  T_{1}^{bn},\hat{B}_{2}-\left(
C_{1}+E_{2}\right)  \right]  ,$ for which the rate $\lambda_{\left(
\mathcal{B},\Delta\right)  }\left(  t\right)  <\varepsilon.$ (That moment
might depend on the parameters $\mathcal{B},\Delta;$ they have however to be
consistent with all the choices already made.)
\end{enumerate}

\noindent Since all the restrictions on $\hat{B}_{2}$ are the lower bounds,
its choice is possible. The time moment $T_{2}^{in}$ is defined to be the
above moment of low input rate. We define $T_{2}^{out}=\hat{B}_{2}-\left(
C_{1}+E_{2}\right)  .$

\subsection{The second induction step}

In this section we will determine the values of the parameters $B_{3}$ and
$d_{2}$ of the server $\mathcal{N}\left(  \mathcal{B},\Delta\right)  .$ We
will do this by looking on the server $\mathcal{N}_{2}=\mathcal{N}\left(
\mathcal{B}_{2},\tilde{\Delta}_{2}\right)  .$ $\mathcal{B}_{2}$ is the set
$\left\{  B_{1}=1,C_{1},B_{2},C_{2}\right\}  ,$ the service time $\eta_{2}$
takes (integer) values in $\left[  1,C_{1}\right]  \cup\left[  B_{2}%
,C_{2}\right]  $ (taking value $1$ with the probability at least $\frac{1}%
{2},$ see $\left(  \ref{13}\right)  $). The value $C_{2}\ $equals to
$B_{2}+G_{2},$ where the parameter $G_{2}$ was fixed earlier. The sequence
$\tilde{\Delta}_{2}$ has two non-zero entries: $\tilde{\Delta}_{2}=\left(
d_{1},1-d_{1}\right)  .$ The initial state is given by the measure
$\nu_{\tilde{\Delta}_{2}},$ see $\left(  \ref{14}\right)  $. The values
$B_{2}$ and $d_{1}$ were chosen earlier.

The server $\mathcal{N}\left(  \mathcal{B}_{2},\tilde{\Delta}_{2}\right)  $
thus belongs to the class of servers $\mathcal{N}\left(  \mathcal{B}%
,\Delta\right)  ,$ considered in the previous section, so all the results of
it apply. In particular, the rate function $\lambda_{\mathcal{B}_{2}%
,\tilde{\Delta}_{2}}$ has the properties that $\lambda_{\mathcal{B}_{2}%
,\tilde{\Delta}_{2}}\left(  T_{1}^{bn}\right)  \approx\Lambda_{\mathcal{B}%
_{1},\tilde{\Delta}_{1}},$ while $\lambda_{\mathcal{B}_{2},\tilde{\Delta}_{2}%
}\left(  T_{2}^{in}\right)  $ is very small.

Still, after some time, longer than $T_{2}^{in},$ the server $\mathcal{N}_{2}
$ will equilibrate, and its state $\mu_{2}^{t}$ will go to its (weak) limit,
$\nu_{2}^{\rho_{2}},$ which is the invariant state of $\mathcal{N}_{2},$ with
$\rho_{2}=N\left(  \mu_{2}^{t}\right)  =N\left(  \nu_{2}^{\rho_{2}}\right)
=N\left(  \nu_{\tilde{\Delta}_{2}}\right)  .$

Let the finite rectangle $I_{2}\subset\Omega_{\left[  0,C_{2}-1\right]  }$ be
the smallest one on which \textquotedblleft almost all of the measure $\nu
_{2}^{\rho_{2}}$ is concentrated\textquotedblright, in the sense of relation
$\left(  \ref{30}\right)  .$ Define the time $T_{2}^{bn}$ as the first moment
$t$ after $T_{2}^{in},$ at which the state $\mu_{2}^{t}$ is $\varepsilon
$-close to $\nu_{2}^{\rho_{2}}.$ Due to the continuity of the dependence of
the restriction $\mu^{T}\Bigm|_{I_{2}}$ on $\mathcal{B},\Delta$ we can claim
the existence of the values $\check{B}_{3}<\infty$ and $\hat{d}_{2}<1-d_{1},$
such that for all $\mathcal{B}$-s and $\Delta$-s with $B_{3}\geq\check{B}_{3}$
and $1-d_{1}>d_{2}\geq\hat{d}_{2}$ the state $\mu^{t}$ of all corresponding
servers $\mathcal{N}\left(  \mathcal{B},\Delta\right)  $ at the moment
$t=T_{2}^{bn}$ are $2\varepsilon$-close to $\nu_{2}^{\rho_{2}}.$

The choice for $d_{2}$ is the value $\hat{d}_{2}.$ Since $d_{1}+d_{2}<1,$
after some long time $t>T_{2}^{bn}$ there will be almost no clients on the
level below $C_{2}$ -- namely, we will have that $N_{\Omega_{\left[
0,C_{2}-1\right]  }}\left(  \mu^{t}\right)  <\varepsilon$ and $\lambda
_{\mathcal{B},\Delta}\left(  t\right)  <\varepsilon$ -- provided only that the
value $B_{3}$ is big enough for the rate $\lambda_{\mathcal{B},\Delta}\left(
t\right)  $ to have time to \textquotedblleft almost\textquotedblright%
\ vanish. (The claim that $N_{\Omega_{\left[  0,C_{2}-1\right]  }}\left(
\mu^{t}\right)  <\varepsilon,$ being valid, is not used in our argument, so we
will omit its proof.) This statement is proven in precisely the same way as in
the previous section, using again the Lemma \ref{decay}. So let $T_{2}%
^{in}>T_{2}^{bn}$ be the first such $t,$ like in the previous section.

\subsection{The end of the proof}

Clearly, the above construction can be iterated. The result of infinite number
of its iterations is the needed server $\mathcal{N}\left(  \mathcal{B}%
,\Delta\right)  .$ The only think to be established is the relation $\left(
\ref{0002}\right)  ,$ together with the positivity of the constant
$c_{PK}\left(  \eta,\rho\right)  $ there.

As was explained earlier, the values of the limiting rate function
$\lambda_{\mathcal{B},\Delta}\left(  t\right)  $ at the moments $t=T_{k}^{bn}%
$, $k=1,2,..$ are very close to the limiting values $\Lambda_{k}\equiv
\Lambda_{\mathcal{B}_{k},\tilde{\Delta}_{k}},$ which are input ($=$output)
rates of the corresponding approximating servers $\mathcal{N}\left(
\mathcal{B}_{k},\tilde{\Delta}_{k}\right)  ,$ as they approach their
stationary states. These limiting rates can be found from the relation, easily
obtained by combination of Little's formula and Pollaczek-Khinchin formula for
$M\Bigm|GI\Bigm|1,\infty$ queuing system:%
\[
\rho_{k}=\frac{\Lambda_{k}^{2}\mathbb{E}\left(  \eta_{k}^{2}\right)
}{2\left(  1-\Lambda_{k}\mathbb{E}\left(  \eta_{k}\right)  \right)  },
\]
see e.g. \cite{St}, sect. 5.0. Here $\eta_{k}$ is the random service time of
the server $\mathcal{N}\left(  \mathcal{B}_{k},\tilde{\Delta}_{k}\right)  ,$
and $\rho_{k}$ is the average number of clients in it. Solving that equation
for $\Lambda_{k}$ we find
\[
\Lambda_{k}=\frac{\sqrt{\left(  \mathbb{E}\left(  \eta_{k}\right)  \rho
_{k}\right)  ^{2}+2\rho_{k}\mathbb{E}\left(  \eta_{k}^{2}\right)  }%
-\mathbb{E}\left(  \eta_{k}\right)  \rho_{k}}{\mathbb{E}\left(  \eta_{k}%
^{2}\right)  }.
\]
By our construction the quantities $\rho_{k},$ $\mathbb{E}\left(  \eta
_{k}\right)  ,$ $\mathbb{E}\left(  \eta_{k}^{2}\right)  $ all go to their
limits as $k\rightarrow\infty,$ which are finite and positive. So the limit%

\[
c_{PK}\left(  \eta,\rho\right)  =\lim_{k\rightarrow\infty}\frac{\sqrt{\left(
\mathbb{E}\left(  \eta_{k}\right)  \rho_{k}\right)  ^{2}+2\rho_{k}%
\mathbb{E}\left(  \eta_{k}^{2}\right)  }-\mathbb{E}\left(  \eta_{k}\right)
\rho_{k}}{\mathbb{E}\left(  \eta_{k}^{2}\right)  }%
\]
is positive as well.

\section{Appendix I. Weak PH for discrete service times}

In this section we will extend the results of \cite{KR}, concerning the Weak
Poisson Hypothesis for the networks on complete graphs, to the situation when
the service time is integer valued. The WPH deals with the following
situation, described already in the introduction. There are $N$ customers,
undergoing service by $M$ servers. The customers are served on the basis of
the First-In-First-Out (FIFO) protocol. Before being served they might wait in
queues. The service times at different servers are i.i.d. random variables,
with the distribution $\eta$. After being served the customer chooses randomly
one of $M$ servers, with probability $\frac{1}{\left|  M\right|  }$ each, and
goes for service to it. So customers never leave the network, nor are there
new customers arrivals into it. This model is suitable for describing real
situations when each client visits many servers before exiting the network.

Once we specify the initial state of the network, its (random) evolution is
defined. In fact, it is Markov process on the set $\bar{\Omega}_{M}$ of
configurations of our network, if by configuration the following is meant:

\begin{itemize}
\item at every node we need to know the length $n$ of the queue,
$n=0,1,2,...,$

\item if $n\geq1$ -- that is, the server is occupied by some client -- we need
to know the time $\tau$ the server already spent working on that client.
\end{itemize}

\noindent A state of the network is therefore a probability measure $\bar{\nu
}$ on $\bar{\Omega}_{M}=\Omega^{M}.$

If $M,N$ are large, the above Markov process becomes quite complicated, and in
order to make it treatable one has to go to the limit $M,N\rightarrow\infty.$
So let us fix sequences $M\rightarrow\infty,$ $N\rightarrow\infty,$ such that
$\frac{N}{M}\rightarrow\rho,$ and suppose that for every $M$ we are given some
initial state $\bar{\nu}_{M}$ on $\bar{\Omega}_{M}.$ To have the convergence
we are looking for, we need of course to impose some conditions on the
sequence $\bar{\nu}_{M}.$ We will need the following properties.

\begin{itemize}
\item \textit{Symmetry. }The permutation group $S_{M}$ acts on $\bar{\Omega
}_{M}$ in a natural way. We want the measures $\bar{\nu}_{M}$ to be $S_{M}$-invariant.

\item \textit{Convergence (LLN). }There is a projection map $\Pi$ from the set
$\bar{\Omega}_{M}$ to the set of probability measures $\mathcal{M}\left(
\Omega\right)  $ on $\Omega,$ which to every point $\bar{\omega}_{M}=\left(
\omega_{1},...,\omega_{M}\right)  \in\bar{\Omega}_{M}$ corresponds the
measure
\[
\Pi\left(  \bar{\omega}_{M}\right)  =\frac{1}{M}\sum_{i=1}^{M}\delta
_{\omega_{i}}.
\]
Therefore the induced map $\Pi:\mathcal{M}\left(  \bar{\Omega}_{M}\right)
\rightarrow\mathcal{M}\left(  \mathcal{M}\left(  \Omega\right)  \right)  $ is
defined. So we have a sequence $\Pi\left(  \bar{\nu}_{M}\right)  $ of the
elements of the set $\mathcal{M}\left(  \mathcal{M}\left(  \Omega\right)
\right)  .$ We need that the week limit $\lim\Pi\left(  \bar{\nu}_{M}\right)
$ exists, and moreover it is not just some element of $\mathcal{M}\left(
\mathcal{M}\left(  \Omega\right)  \right)  ,$ but is atomic, i.e. belongs to
the image of the natural inclusion of $\mathcal{M}\left(  \Omega\right)  $ in
$\mathcal{M}\left(  \mathcal{M}\left(  \Omega\right)  \right)  .$ Namely, this
inclusion assigns to every measure $\nu$ on $\Omega$ the measure $\delta_{\nu
}$ on $\mathcal{M}\left(  \Omega\right)  .$ So we need that
\begin{equation}
\lim\Pi\left(  \bar{\nu}_{M}\right)  =\delta_{\nu} \label{0031}%
\end{equation}
for some $\nu.$ In other words, we need the Law of Large Numbers to hold for
the sequence $\Pi\left(  \bar{\nu}_{M}\right)  .$
\end{itemize}

\begin{theorem}
\label{WPH} Let $\bar{\mu}_{M}^{t}$ be the state of the above defined Markov
process on $\bar{\Omega}_{M}$ at the time $t,$ with symmetric initial state
$\bar{\mu}_{M}^{0}=\bar{\nu}_{M}.$ Let us denote by $\bar{\mu}_{M|K}^{t}$ its
restriction to $\bar{\Omega}_{K}.$

Suppose that $\left(  \ref{0031}\right)  $ holds. Then for every $K\geq1$ the
limiting projection $\bar{\mu}_{\infty|K}^{t}=\lim_{M\rightarrow\infty}%
\bar{\mu}_{M|K}^{t}$ splits into a product of $K$ independent identical
processes, $\mu^{t},$ which are NMP-s with initial state $\mu^{0}=\nu.$ (For
$\eta$ taking integer values the process $\mu^{t}$ is defined in the Section 2.)
\end{theorem}

For continuous service time $\eta$ the corresponding process $\mu^{t}$ is
described in details in \cite{KR} or \cite{RS}.

In the continuous case the proof of the above theorem is the subject of the
paper \cite{KR} and is quite involved. The idea is to study the generators
$\Delta_{M}$ of the processes $\bar{\mu}_{M}^{t}$ and to consider their limit,
$\Delta.$ Then one has to build the process corresponding to $\Delta$ and to
use the Trotter-Kurtz Theorem, which gives conditions of convergence of Markov
processes to the limiting process in terms of convergence of the generators.

In the discrete case the situation is much simpler.

We begin with a simple lemma.

\begin{lemma}
Let the measures $\bar{\nu}_{M}$ satisfy the symmetry condition. Then the
convergence condition is equivalent to the Propagation of Chaos property: for
every $K\geq1$
\begin{equation}
\lim_{M\rightarrow\infty}\bar{\nu}_{M|K}=\underset{K}{\underbrace{\nu
\times...\times\nu}}. \label{0040}%
\end{equation}
In fact, it is sufficient to consider only the $K=2$ case.
\end{lemma}

\begin{proof}
By de Finetti theorem we know that the limit measures $\lim_{M\rightarrow
\infty}\bar{\nu}_{M|K}$ is either a product measure, like in $\left(
\ref{0040}\right)  ,$ or a mixture of such products, $\int\underset
{K}{\underbrace{\varkappa\times...\times\varkappa}}~\mathfrak{d}\left(
d\varkappa\right)  ,$ where the measure $\mathfrak{d}\in\mathcal{M}\left(
\mathcal{M}\left(  \Omega\right)  \right)  $ is non-atomic and does not depend
on $K.$ But in the latter case we have that $\lim_{M\rightarrow\infty}%
\Pi\left(  \bar{\nu}_{M}\right)  =\mathfrak{d}.$

Conversely, if the limit $\lim_{M\rightarrow\infty}\Pi\left(  \bar{\nu}%
_{M}\right)  =\mathfrak{d}$ is non-atomic, then the two components of the
measure $\bar{\nu}_{M|K=2}$ are dependent even in the limit $M\rightarrow
\infty.$
\end{proof}

\medskip

\noindent\textbf{Proof of the Theorem 3. }Our proof goes by induction in $t$.
Clearly, we will be done once we prove the statement of our theorem for just
one value of time, $t=1.$

Let us first investigate the total flow $\Sigma_{M}$ of the customers from all
the $M$ servers after the first step. If we condition on the initial
configuration $\bar{\omega}_{M}=\left(  \omega_{1}=\left(  n_{1},\tau
_{1}\right)  ,...,\omega_{M}=\left(  n_{M},\tau_{M}\right)  \right)  \in
\bar{\Omega}_{M}$, then the total flow is given by the sum $\Sigma
_{M}\Bigm|\bar{\omega}_{M}=\sigma_{1}+...+\sigma_{M},$ where $\sigma_{i}$ are
independent random variables distributed as follows:
\[
\sigma_{i}=\left\{
\begin{array}
[c]{ll}%
1 & \text{ with probability }p_{\tau_{i}}\text{,}\\
0 & \text{ with probability }1-p_{\tau_{i}},
\end{array}
\right.
\]
see $\left(  \ref{32}\right)  .$ The unconditional flow is then given by
\[
\Sigma_{M}=\sum_{\bar{\omega}_{M}\in\bar{\Omega}_{M}}\left(  \Sigma
_{M}\Bigm|\bar{\omega}_{M}\right)  \bar{\nu}_{M}\left(  \bar{\omega}%
_{M}\right)  .
\]
We need first to establish the Law of Large Numbers for the sequence
$\Sigma_{M}.$ To that end let us denote by $C_{\tau}\left(  \bar{\omega}%
_{M}\right)  $ the number of indexes $i\in\left\{  1,2,...,M\right\}  ,$ for
which $\tau_{i}=\tau.$ Then
\[
\Sigma_{M}\Bigm|\bar{\omega}_{M}=\sum_{\tau=1}^{\infty}\sum_{j=1}^{C_{\tau
}\left(  \bar{\omega}_{M}\right)  }\sigma_{j}^{\tau},
\]
where $\left\{  \sigma_{j}^{\tau}:j,\tau=1,2,...\right\}  $ are independent
random variables, and for every fixed $\tau$ $\left\{  \sigma_{j}^{\tau
}:j=1,2,...\right\}  $ are i.i.d., with
\[
\sigma_{j}^{\tau}=\left\{
\begin{array}
[c]{ll}%
1 & \text{ with probability }p_{\tau}\text{,}\\
0 & \text{ with probability }1-p_{\tau}.
\end{array}
\right.
\]
Clearly, LLN holds for $\Sigma_{M}$ once it holds for random variables
$C_{\tau}\left(  \bar{\omega}_{M}\right)  ,$ which distribution is governed by
the measures $\bar{\nu}_{M}.$ But the convergence in probability
$\frac{C_{\tau}\left(  \bar{\omega}_{M}\right)  }{M}\rightarrow\sum_{n}%
\nu\left(  n,\tau\right)  $ follows immediately from LLN for the sequence
$\Pi\left(  \bar{\nu}_{M}\right)  .$

By symmetry, the average flow to any given server and from any given server in
the state $\bar{\mu}_{M}^{t}$ are the same. So we need only to check that the
flow to any given server is asymptotically Poisson, and that the flows to any
two given servers are asymptotically independent.

The number of customers $\mathfrak{n}^{M}$, going to a given -- say, first --
server, is given by
\[
\mathfrak{n}^{M}=\sum_{i=1}^{\Sigma_{M}}\xi_{i},
\]
where the random variables $\xi_{i}$ are i.i.d.,
\begin{equation}
\xi_{i}=\left\{
\begin{array}
[c]{ll}%
1 & \text{ with probability }\frac{1}{M}\text{,}\\
0 & \text{ with probability }1-\frac{1}{M}.
\end{array}
\right.  \label{0310}%
\end{equation}
The meaning of the event $\xi_{i}=1$ is that the $i$-th customer,
$i=1,...,\Sigma_{M},$ decides to go to the first server for the next service.
But since the random variable $\Sigma_{M}$ satisfies LLN with the mean value
$\sim M$ -- in fact,
\[
\frac{\mathbb{E}\left(  \Sigma_{M}\right)  }{M}\rightarrow\int_{\Omega}%
p_{\tau}~d\nu
\]
-- the distribution of $\mathfrak{n}^{M}$ is asymptotically Poisson, due to
the Poisson limit theorem.

The flow of customers $\left(  \mathfrak{n}_{1}^{M},\mathfrak{n}_{2}%
^{M}\right)  ,$ going to the first two servers, is the random vector
\[
\left(  \mathfrak{n}_{1}^{M},\mathfrak{n}_{2}^{M}\right)  =\sum_{i=1}%
^{\Sigma_{M}}\zeta_{i},
\]
where
\[
\zeta_{i}=\left\{
\begin{array}
[c]{ll}%
\left(  1,0\right)  & \text{ with probability }\frac{1}{M}\text{,}\\
\left(  1,0\right)  & \text{ with probability }\frac{1}{M}\text{,}\\
\left(  0,0\right)  & \text{ with probability }1-\frac{2}{M}\text{.}%
\end{array}
\right.
\]
Clearly, the random variables $\mathfrak{n}_{1}^{M},\mathfrak{n}_{2}^{M}$ are
not independent. They are, however, independent asymptotically, when
$M\rightarrow\infty.$ To see it, consider the conditional distribution of
$\mathfrak{n}_{1}^{M}$ under condition that $\mathfrak{n}_{2}^{M}=k.$ We have
\[
\mathfrak{n}_{1}^{M}\Bigm|_{\mathfrak{n}_{2}^{M}=k}=\sum_{i=1}^{\Sigma_{M}%
-k}\tilde{\xi}_{i},
\]
with $\tilde{\xi}_{i}$ being given by
\[
\tilde{\xi}_{i}=\left\{
\begin{array}
[c]{ll}%
1 & \text{ with probability }\frac{\frac{1}{M}}{\left(  1-\frac{1}{M}\right)
}\text{,}\\
0 & \text{ with probability }\frac{1-\frac{2}{M}}{1-\frac{1}{M}},
\end{array}
\right.
\]
so Poisson limit theorem still applies. But since for every $k$%
\[
\lim_{M\rightarrow\infty}\frac{\frac{1}{M}}{\left(  1-\frac{1}{M}\right)
}\left(  \mathbb{E}\left(  \Sigma_{M}\right)  -k\right)  =\lim_{M\rightarrow
\infty}\frac{1}{M}\mathbb{E}\left(  \Sigma_{M}\right)  ,
\]
the conditional distribution $\mathfrak{n}_{1}^{M}\Bigm|_{\mathfrak{n}_{2}%
^{M}=k}$ converges to that of $\mathfrak{n}_{1}^{M}.$ This convergence is of
course not uniform in $k,$ but since the tail of the Poisson distribution
decays exponential, it is immaterial. $\blacksquare$

\section{Appendix II. Full PH for discrete service times}

As is explained in \cite{RS}, we have to consider the following random
processes, associated with a single server: given any function $\lambda\left(
t\right)  \geq0,$ $t=0,1,2,...,$ we consider the corresponding integer Poisson
process of arriving customers. It is defined in the standard way: the
probability that $n$ customers arrive at a given moment $t\in\mathbb{N}$ is
given by
\[
e^{-\lambda\left(  t\right)  }\frac{\lambda\left(  t\right)  ^{n}}{n!},
\]
while the arrivals at different moments are independent. The customers are
served in the order they come, and the service times are given by the
independent realizations of a random variable $\eta,$ which in our case will
take only finitely many integer values. We suppose that the (finite) support
$\mathfrak{S}\left(  \eta\right)  \subset\mathbb{N}$ of the distribution
function $p$ of $\eta$ contains the value $1\in\mathbb{N},$ i.e. $p\left(
1\right)  >0.$ In particular, the case of $\eta\equiv1,$ considered earlier by
Stolyar, see \cite{S}, is included. We denote by $\mathcal{T}$ the maximal
element in $\mathfrak{S}\left(  \eta\right)  ,$ and we put $E=\mathbb{E}%
\left(  \eta\right)  .$

A configuration on the server at a given time moment $t$ consists of the
number $n\geq0$ of customers waiting to be served plus the duration $\tau$ of
the elapsed service time of the customer under the service at the moment $t.$
The set of configurations $\Omega$ is thus the set of all pairs $\left(
n,\tau\right)  ,$ with integers $n>0$ and $\tau\in\lbrack0,\mathcal{T}-1],$
plus the point $\mathbf{0}$, describing the situation of the server being
idle. For a configuration $\omega=\left(  n,\tau\right)  \in\Omega$ we define
$N\left(  \omega\right)  =n.$ We put $N\left(  \mathbf{0}\right)  =0.$ A state
of the server is any probability measure on $\Omega.$

Once the initial state $\nu$ and the rate function $\lambda\left(  t\right)  $
are given, the evolving state $\mu_{\nu,\lambda}\left(  t\right)  $ is
defined, $\mu_{\nu,\lambda}\left(  0\right)  =\nu,$ which constitutes a
non-homogeneous Markov process -- General Flow Process (GFP) in terminology of
\cite{RS}. In particular, the exit flow of customers whose service is over, is
defined. Let $b\left(  t\right)  $ be the mean number of the customers exiting
the server at the moment $t.$ The function $b$ is uniquely defined by the
function $\lambda$ and initial state $\nu;$ therefore we can write that
$b\left(  \cdot\right)  =A\left(  \nu,\lambda\left(  \cdot\right)  \right)  ,$
where $A$ is the corresponding operator.

For constant $\lambda\equiv c$ the process becomes homogeneous, and we have
the weak convergence $\mu_{\nu,\lambda}\left(  t\right)  \rightarrow\nu_{c}$
as $t\rightarrow\infty,$ for any initial state $\nu.$

The Non-Linear Markov Process (NMP) is a special case of GFP, which
corresponds to the function $\lambda_{\nu},$ satisfying the fix-point
equation
\[
\lambda_{\nu}\left(  \cdot\right)  =A\left(  \nu,\lambda_{\nu}\left(
\cdot\right)  \right)  .
\]
The above notation suggests that for every initial state $\nu$ the
corresponding NMP exists and is unique; this is indeed the case, see \cite{RS}
and \cite{KR}.

As in \cite{RS}, the validity of the Poisson Hypothesis in our setting holds
once the following is proven:

\begin{theorem}
\label{phd} For every initial state $\nu$ with finite mean queue:
\[
N\left(  \nu\right)  \equiv\mathbb{E}_{\nu}\left(  N\left(  \omega\right)
\right)  <\infty,
\]
the solution $\lambda_{\nu}\left(  \cdot\right)  $ of the equation
\[
A\left(  \nu,\lambda\left(  \cdot\right)  \right)  =\lambda\left(
\cdot\right)
\]
has the \textbf{relaxation }property:
\[
\lambda_{\nu}\left(  t\right)  \rightarrow c\text{ as }t\rightarrow\infty,
\]
where the constant $c$ satisfies for every $t$ the relation
\[
N\left(  \mu_{\nu,\lambda}\left(  t\right)  \right)  \equiv\mathbb{E}%
_{\mu_{\nu,\lambda}\left(  t\right)  }\left(  N\left(  \omega\right)  \right)
=N\left(  \nu_{c}\right)  \equiv\mathbb{E}_{\nu_{c}}\left(  N\left(
\omega\right)  \right)  .
\]
Moreover, $\mu_{\nu,\lambda_{\nu}\left(  \cdot\right)  }\left(  t\right)
\rightarrow\nu_{c}$ weakly, as $t\rightarrow\infty.$
\end{theorem}

In the following we are indicating the changes in the proof of the main result
of \cite{RS} -- \textit{Theorem 1} -- needed in order to extend it to the case
treated by the above theorem. Here and later we will use italics to indicate
the statements in \cite{RS}.

The first statements -- from \textit{Theorem 3} to \textit{Theorem 6} --
remain unchanged. In particular, we have the validity of the key
self-averaging relation:
\[
b\left(  x\right)  =\left[  \lambda\ast q_{\lambda,x}\right]  \left(
x\right)
\]
for some probability kernels $q_{\lambda,x}$; the only difference now is that
all the measures $q_{\ast,\ast}$ have integer supports. To see that the proofs
of these statements can be extended to cover the new setting, one can argue as
follows: let us approximate weakly our (atomic) rate measure $\sum
_{n\in\mathbb{N}}\lambda\left(  n\right)  \delta_{n}$ by the continuous rate
measure $\lambda^{\varepsilon}\left(  t\right)  dt,$ where
\[
\lambda^{\varepsilon}\left(  t\right)  =\left\{
\begin{array}
[c]{ll}%
\frac{\lambda\left(  n\right)  }{2\varepsilon} & \text{ if }\left\vert
t-n\right\vert \leq\varepsilon,\\
0 & \text{ otherwice.}%
\end{array}
\right.
\]
We can also approximate weakly the discrete probability distribution $p\left(
n\right)  $ by a density $p^{\varepsilon}\left(  t\right)  >0.$ The results of
\textit{Theorems 3-6} are applicable to the $\varepsilon$-approximations.
Since the statements are just identities, we can pass then to the limit
$\varepsilon\rightarrow0.$

Let us go to \textit{Lemma 7, }which is based on \textit{Proposition 8,} which
in turn follows from \textit{Lemmas 11 \& 12. }The statement of \textit{Lemma
7 }-- relation \textit{(55) }-- should now read:
\[
\sum_{t=s}^{s+T}\lambda_{\nu}\left(  t\right)  \,<T\left(  E-\varepsilon
^{\prime}\right)  :
\]
the integral is replaced by the sum, while instead of $1$ we need to use $E$
-- the mean value of $\eta$ (the mean one assumption is no loss of generality
for the real valued random variables, but not for integer valued ones!).
\textit{Lemma 11} deals with the Poisson flows with continuous rate functions
$\lambda\left(  t\right)  .$ Two proofs of that lemma are given in \cite{RS},
and the second one is using the discrete approximations, followed by the limit
procedure. But the discrete approximation step itself gives the proof of the
modified statement we need, applicable to integer Poisson process of customer arrivals.

The following statement is the replacement of \textit{Lemma 12}; though the
changes are infinitesimal, we present here this discrete counterpart of this
lemma, in order to dispel even the slightest problems the reader might have.

In what follows we denote by $\kappa^{\left(  a\right)  }$ the measure on
$\left[  A,B\right]  \subset\mathbb{R}^{1}$, having the constant density
$\lambda\left(  t\right)  =a.$ We denote by $\bar{\kappa}^{\left(  a\right)
}$ the atomic measure counterpart on $\left[  A,B\right]  \subset\mathbb{N}:$
for any integer $t\in\left[  A,B\right]  ,$ $\bar{\kappa}^{\left(  a\right)
}\left(  t\right)  =a$.

\begin{lemma}
\label{calculdiscreet} Let the atomic measure $\chi$ on $\left[  A,B\right]
\equiv\left\{  A,A+1,...,B\right\}  \subset\mathbb{N}$ is bounded, $0\leq
\chi\left(  t\right)  \leq L$ for every $t\in\left[  A,B\right]  ,$ and
satisfies the property:
\[
\chi\left(  \left[  A,B\right]  \right)  \geq\left(  E-\varepsilon\right)
\left(  B+1-A\right)  .
\]
Then there exists a segment $\left[  A,C\right]  \subset\left[  A,B\right]  $
of the length
\begin{equation}
C+1-A>\frac{\varepsilon}{L}\left(  B+1-A\right)  , \label{305}%
\end{equation}
such that
\[
\chi\Bigm|_{\left[  A,C\right]  }\succ\bar{\kappa}^{\left(  E-2\varepsilon
\right)  }\Bigm|_{\left[  A+1,C-1\right]  }.
\]

\end{lemma}

\begin{proof}
Let us fix some number $M>2,$ and consider the measure $\chi^{M}$ on $\left[
A-\frac{1}{2},B+\frac{1}{2}\right]  \subset\mathbb{R}^{1},$ defined by the
density
\[
\lambda^{M}\left(  t\right)  =\left\{
\begin{array}
[c]{ll}%
M\chi\left(  n\right)  & \text{ if }\left\vert t-n\right\vert <\frac{1}%
{2M}\text{ for some integer }n\in\left[  A,B\right]  ,\\
0 & \text{ otherwice.}%
\end{array}
\right.
\]
Let us smooth the function $\lambda^{M}\left(  t\right)  $ by a small
perturbation; the smoothed version of it still will be denoted by $\lambda
^{M}\left(  t\right)  .$ We have that the measure $\chi^{M}$ on $\left[
A-\frac{1}{2},B+\frac{1}{2}\right]  $ satisfies $\chi^{M}\left(  \left[
A-\frac{1}{2},B+\frac{1}{2}\right]  \right)  \geq\left(  E-\varepsilon\right)
\left(  B+\frac{1}{2}-\left(  A-\frac{1}{2}\right)  \right)  ,$ while
$0<\lambda^{M}\left(  t\right)  \leq ML.$ We can well repeat for the measure
$\chi^{M}$ the steps of the proof of \textit{Lemma 12. }As there, we consider
the disjoint maximal segments $\left[  C_{1},D_{1}\right]  ,\left[
C_{2},D_{2}\right]  ,...\subset\left[  A-\frac{1}{2},B+\frac{1}{2}\right]  $
of the family of all segments $\left[  c,d\right]  \subset\left[  A-\frac
{1}{2},B+\frac{1}{2}\right]  ,$ for which we have the property
\begin{equation}
\chi^{M}\Bigm|_{\left[  c,d\right]  }\succ\kappa^{\left(  E-2\varepsilon
\right)  }\Bigm|_{\left[  c,d\right]  }, \label{0309}%
\end{equation}
and denote by $\left[  A-\frac{1}{2},a\right]  $ the segment $\left[
C_{i},D_{i}\right]  $ among these maximal, which contains the point
$A-\frac{1}{2}$. For all the segments $\left[  C_{k},D_{k}\right]  $ except
$\left[  A-\frac{1}{2},a\right]  ,$ we again have
\begin{equation}
\chi^{M}\left(  \left[  C_{k},D_{k}\right]  \right)  =\left(  E-2\varepsilon
\right)  \left(  D_{k}-C_{k}\right)  , \label{0027}%
\end{equation}
while for any point $x\in\left[  A-\frac{1}{2},B+\frac{1}{2}\right]  $ outside
all of the segments $\left[  C_{i},D_{i}\right]  ,$ we have $\lambda
^{M}\left(  x\right)  \leq E-2\varepsilon.$ Together with $\left(
\ref{0027}\right)  $ it implies that
\begin{equation}
\chi^{M}\left(  \left[  a,B+\frac{1}{2}\right]  \right)  \leq\left(
E-2\varepsilon\right)  \left(  B+\frac{1}{2}-a\right)  . \label{0028}%
\end{equation}
On the other hand,
\begin{align}
\chi^{M}\left(  \left[  A-\frac{1}{2},B+\frac{1}{2}\right]  \right)   &
=\chi^{M}\left(  \left[  A-\frac{1}{2},a\right]  \right)  +\chi\left(  \left[
a,B+\frac{1}{2}\right]  \right) \label{0029}\\
&  \geq\left(  E-\varepsilon\right)  \left(  B+1-A\right)  .\nonumber
\end{align}
Therefore the segment $\left[  A-\frac{1}{2},a\right]  $ is non-trivial. By
definition $\left(  \ref{0309}\right)  $ we have that $\lambda^{M}\left(
a\right)  \geq E-2\varepsilon.$ From the maximality of the segment $\left[
A-\frac{1}{2},a\right]  $ it then follows that $a=m+\frac{1}{2M}$ for some
integer $m\in\left[  A,B\right]  .$ Clearly, $\chi^{M}\left(  \left[
A-\frac{1}{2},m+\frac{1}{2M}\right]  \right)  \leq L\left(  m+1-A\right)  ,$
while $\left(  \ref{0028}\right)  $ reads
\[
\chi^{M}\left(  \left[  m+1-\frac{1}{2M},B+\frac{1}{2}\right]  \right)
\leq\left(  E-2\varepsilon\right)  \left(  B+\frac{1}{2}-\left(  m+1-\frac
{1}{2M}\right)  \right)  .
\]
Together with $\left(  \ref{0029}\right)  $ it implies that
\begin{equation}
m+1-A\geq\frac{\varepsilon}{L}\left(  B+1-A\right)  . \label{0030}%
\end{equation}
So we have
\[
\chi^{M}\Bigm|_{\left[  A-\frac{1}{2},m+\frac{1}{2M}\right]  }\succ
\kappa^{\left(  E-2\varepsilon\right)  }\Bigm|_{\left[  A-\frac{1}{2}%
,m+\frac{1}{2M}\right]  }.
\]
Comparison with the atomic counterpart --
\[
\kappa^{\left(  E-2\varepsilon\right)  }\Bigm|_{\left[  A-\frac{1}{2}%
,m+\frac{1}{2M}\right]  }\succ\bar{\kappa}^{\left(  E-2\varepsilon\right)
}\Bigm|_{\left[  A+1,m-1\right]  }%
\]
-- gives us the relation
\[
\chi^{M}\Bigm|_{\left[  A-\frac{1}{2},m+\frac{1}{2M}\right]  }\succ\bar
{\kappa}^{\left(  E-2\varepsilon\right)  }\Bigm|_{\left[  A+1,m-1\right]  }.
\]
Since it holds for every $M,$ we conclude that
\[
\chi\Bigm|_{\left[  A,m\right]  }\succ\bar{\kappa}^{\left(  E-2\varepsilon
\right)  }\Bigm|_{\left[  A+1,m-1\right]  }.
\]
From $\left(  \ref{0030}\right)  $ the proof now follows.
\end{proof}

\textit{Lemma 13 }remains unchanged. \textit{Lemma 14, }stating the Lipschitz
property of the function $\lambda,$ cannot be valid, of course, in our case.
In fact, in the discrete case no analog of Lipschitz property is needed, as we
will explain later.

To get the counterpart to the \textit{Lemma 15, }which provides a lower bound
for the kernels $q_{\ast,\ast},$ we first establish a lower bound on the rates
$\lambda_{\nu}\left(  t\right)  $ of the NMP-s, $t\in\mathbb{N}$.

\begin{lemma}
\label{lowerbound1} Let $\nu$ be any initial state, i.e. a probability measure
on $\Omega,$ and let $\mu_{\nu,\lambda_{\nu}\left(  \cdot\right)  }\left(
t\right)  $ be \textit{the corresponding non-linear Markov process. Then there
exist a time duration }$\mathcal{T}$ and a constant $c>0,$ both depending only
on $\eta,$ such that
\[
\lambda_{\nu}\left(  t\right)  \geq c\text{ provided }t\geq\mathcal{T}.
\]

\end{lemma}

\begin{proof}
The statement is almost evident. For example, the value $\mathcal{T}%
=\max\left\{  k:k\in\mathfrak{S}\left(  \eta\right)  \right\}  $ will go.
Indeed, during any $\mathcal{T}$ consecutive time moments
$t+1,t+2,...,t+\mathcal{T}$ there has to be at least one, $t^{0},$ at which
the mean exit number of customers -- i.e. the value $\lambda\left(
t_{0}\right)  $ -- is at least $\frac{1}{\mathcal{T}}.$ The next such moment
$t^{1}$ will happen not later than $t^{0}+$ $\mathcal{T}.$ But at the moment
$t_{0}$ the average number of arriving customers is also $\lambda\left(
t_{0}\right)  ,$ and since each of them has a positive probability $p\left(
1\right)  $ of finishing his service in the unit time, all the values
$\lambda\left(  t_{0}+1\right)  ,\lambda\left(  t_{0}+2\right)  ,...,\lambda
\left(  t_{0}+\mathcal{T}\right)  $ are bounded from below by a constant $c,$
depending only on $\mathcal{T}$ and $p\left(  1\right)  .$
\end{proof}

The replacement to \textit{Lemma 15 }can now be presented.

\begin{lemma}
\label{lowerb} Let the function $\lambda$ satisfies
\begin{equation}
C\geq\lambda\geq c>0. \label{306}%
\end{equation}
Then
\begin{equation}
q_{\lambda,y}\left(  t\right)  \geq\mathbf{\Pr}\left\{  \text{server is idle
at the moment }y-t\right\}  e^{-C}\frac{c^{t-1}}{\left(  t-1\right)
!}p\left(  1\right)  ^{t}. \label{301}%
\end{equation}

\end{lemma}

For the case of NMP the condition \ref{306} holds, according to Lemma
\ref{lowerbound1}.

\begin{proof}
The relation $\left(  \ref{301}\right)  $ follows easily from the definition,
see relations \textit{(27), (28) }in \cite{RS}. We already have a lower
estimate for the first factor. Let us consider the second one, i.e. the
conditional probability
\[
\mathbf{\Pr}\left\{  \left.
\begin{array}
[c]{c}%
\text{the server is never idle during }\\
\left[  u,u+t\right]  ;\text{ at }u+t\text{ }\\
\text{some client leaves the server }%
\end{array}
\right\vert
\begin{array}
[c]{l}%
\text{the server is idle}\\
\text{ just before }u,\\
\text{at }u\text{ a client arrives}%
\end{array}
\right\}  .
\]
Clearly, the event needed will happen, if at moment $u$ at least $t-1$ more
clients will arrive, while the first $t$ clients will have their service time
equal to $1$. The probability of the former event is $p\left(  1\right)  ^{t}%
$. The probability of arriving $t-1$ extra customers is at least
\[
e^{-C}\frac{c^{t-1}}{\left(  t-1\right)  !}.
\]

\end{proof}

\textit{Lemmas 16-18} and\textit{\ Theorem 19} do not need any alterations.

Next ingredient needed for the proof of \textit{Theorem 1} -- or our Theorem
\ref{phd}\textit{\ -- }is \textit{Lemma 23}. Its discrete analog states in
particular that if the equation
\begin{equation}
f\left(  x\right)  =\sum_{y=0}^{x}f\left(  x-y\right)  q_{x}\left(  y\right)
\label{0211}%
\end{equation}
holds, then under certain assumptions on the probability kernels $q$ the
function $f$ has to flatten out:

Let $M=\limsup_{x\rightarrow\infty}f\left(  x\right)  .$ Then for every $T$
and every $\varepsilon$ given there exists some value $K_{1},$ such that
\begin{equation}
\min_{x\in\left[  K_{1},K_{1}+T\right]  }f\left(  x\right)  \geq
M-\varepsilon. \label{0001}%
\end{equation}
The properties of $q$ needed were:

\begin{itemize}
\item compactness \textit{(99)}: for every $\varepsilon>0$ there exists a
value $K\left(  \varepsilon\right)  ,$ such that
\begin{equation}
\sum_{0}^{K\left(  \varepsilon\right)  }q_{x}\left(  y\right)  \,\geq
1-\varepsilon\label{0111}%
\end{equation}
uniformly in $x;$

\item lower bound \textit{(100)}: for every $T$ the quantity
\[
F_{T}=\inf_{x\geq X\left(  T\right)  }\inf_{y\in\left\{  1,2,...,T\right\}
}q_{x}\left(  y\right)  \,
\]
is positive for some choice of the function $X\left(  T\right)  <\infty
;$\textit{\ }

\item the solution $f$ of $\left(  \ref{0211}\right)  $ is bounded from above,
$f\leq C,$ and has Lipschitz property.
\end{itemize}

We have already commented on the validity of the first two in our case; the
Lipschitz property will not be needed, as we are going to show now.

\textbf{Proof of }$\left(  \ref{0001}\right)  \mathbf{.}$ Let $\delta>0.$ Then
there exists a value $S=S\left(  \delta\right)  >0,$ such that for all $x>S$
we have $f\left(  x\right)  <M+\delta.$ Further, there exists a value
$R=R\left(  \delta\right)  >S,$ such that for all $y\geq R$
\[
\sum_{x=R-S}^{\infty}q_{y}\left(  x\right)  <\delta,
\]
see (\ref{0111}). Finally, there exists a point $y=y\left(  \delta\right)
>R+T,$ such that $f\left(  y\right)  >M-\delta.$ Due to the equation $\left(
\ref{0211}\right)  $ we have for this point
\[
f\left(  y\right)  =\sum_{t=0}^{y-S}f\left(  y-t\right)  q_{y}\left(
t\right)  \,+\sum_{t=y-S+1}^{\infty}f\left(  y-t\right)  q_{y}\left(
t\right)  \,.
\]
Let $A=\left\{  x\in\left[  y-T,y\right]  :f\left(  x\right)  <M-\varepsilon
\right\}  ,$ while $a=\sum_{t\in A}q_{y}\left(  t\right)  \,.$ We want to show
that $a$ has to be small for small $\delta$. Splitting the first sum into two,
according to whether the point $y-t$ is inside $A$ or outside, we have
\[
M-\delta<a\left(  M-\varepsilon\right)  +\left(  1-a\right)  \left(
M+\delta\right)  +C\delta,
\]
which implies that
\[
a<\frac{\left(  2+C\right)  \delta}{\varepsilon+\delta}.
\]
The last ratio goes to zero with $\delta$ for every fixed $\varepsilon$. In
particular, we can choose $\delta=\delta\left(  \varepsilon\right)  $ so small
that $\frac{\left(  2+C\right)  \delta}{\varepsilon+\delta}<F_{T}.$ That
however is consistent with the definition of $F_{T}$ only if $A=\varnothing.$
$\blacksquare$

The rest of the statements of \cite{RS}, needed to obtain the proof of Theorem
\ref{phd}, does not need any modifications. The only exception is the
statement of \textit{Lemma 31, }where one has to replace the bound $\bar
{c}\left(  \nu\right)  <1$ by $\bar{c}\left(  \nu\right)  <E.$

\end{document}